\documentclass[lettersize,journal]{IEEEtran}
\usepackage{amsmath,amsfonts}
\usepackage{array}
\usepackage{textcomp}
\usepackage[caption=false,font=footnotesize,labelfont=sf,textfont=sf]{subfig}
\usepackage{stfloats}
\usepackage{url}
\usepackage{cite}
\usepackage{xcolor}
\usepackage{hyperref}
\usepackage{tabularx}
\usepackage{multirow}
\usepackage{makecell}
\usepackage{graphicx}

\begin{document}

\title{Thinking in Tokens, Talking in Bits: \\ A Practical Interface for Token Communication}

\author{ Chanho Park, \IEEEmembership{Member,~IEEE,} Bumsu Park, \IEEEmembership{Student Member,~IEEE,} \\
Soonhee Kwon, Sangrim Lee, \IEEEmembership{Member,~IEEE,} and Namyoon Lee, \IEEEmembership{Senior Member,~IEEE}
\thanks{ Chanho Park, Bumsu Park, and Namyoon Lee are with the Department of Electrical Engineering, POSTECH, Pohang 37673, South Korea (e-mail: \{ chanho26, bumsupark, nylee\}@postech.ac.kr).

Soonhee Kwon and Sangrim Lee are with the Communication \& Media Standard Lab., LG Electronics, Seoul 06772, South Korea (e-mail: \{soonhee.kwon, sangrim.lee\}@lge.com).}
}

\markboth{Submitted to IEEE Communications Magazine}%
{Park \MakeLowercase{\textit{et al.}}: Thinking in Tokens, Talking in Bits}

\maketitle

\begin{abstract}
Advanced artificial intelligence models think in tokens; contemporary communication systems carry bits. The direct way to bridge this gap is to transmit tokens, but that makes a model-specific representation part of the air interface, coupling the endpoints through a shared tokenizer, codebook, and often a neural transceiver. We take a different route: keep bits in the payload and let tokens control how those bits are generated and protected. The resulting \emph{token-bit interface transition} aligns task-side tokens with source- and channel-coding units, translates token relevance into codec controls, and preserves the induced priority order across the coding chain. We instantiate it for image classification, where a vision transformer scores the task relevance of each image region from its attention maps: those scores steer block-wise JPEG rate allocation, then group the compressed bits for protection at different polar-code rates. The payload remains an explicit, reconstructable bitstream recovered by a correspondingly configured decoder. Over-the-air experiments on a software-defined radio testbed show improved accuracy--latency tradeoffs over separate source-channel coding, performance competitive with far more memory-intensive neural joint source-channel coding, and graceful degradation under channel mismatch. Token communication, then, need not transmit tokens explicitly; what it needs is an interface through which tokens determine how bits are communicated.
\end{abstract}


\section{Introduction}
\IEEEPARstart{F}{uture} wireless networks will carry large volumes of multimodal data for real-time artificial intelligence (AI) services such as cooperative perception and connected autonomy. In most of these services, the receiver does not want the source; it wants the answer the source supports---a label, a decision, a maneuver. This is the premise of {\it semantic communication}, which allocates resources according to what the downstream task needs rather than what exact reconstruction demands \cite{gunduz2023beyond}. Modern AI models suggest a natural unit for doing so: they decompose text, images, audio, and video into tokens, and their attention over those tokens reveals, almost incidentally, which ones matter for the decision at hand. {\it Token communication} elevates these tokens to the organizing unit of task-oriented transmission \cite{devoto2026adaptive}.

The open question is how a communication system should act on token-level relevance. The dominant answer has been to learn the entire transceiver: deep joint source-channel coding (DeepJSCC) maps the source directly to channel symbols, optimizing end-to-end performance and, in task-oriented variants, downstream accuracy \cite{bourtsoulatze2019deep, lyu2024semantic}. Performance is strong, but the prioritization is implicit---buried in learned weights---and the analog interface it produces does not conform to standardized digital pipelines. Digital alternatives instead retain separate source-channel coding (SSCC) structures while injecting learned representations or relevance signals into selected blocks \cite{huang2025d2}. Compatibility improves, but the injections remain block-local: no general interface has emerged for deriving codec-specific controls from token-level relevance and carrying them across the whole coding chain.

A third line transmits the tokens themselves---discrete token sequences or token-derived neural representations as payload \cite{devoto2026adaptive, qiao2025token}. This aligns what is transmitted with what AI models process, but at a price. Both endpoints must maintain compatible tokenizers, codebooks, and model versions, and conveying token embeddings as channel symbols requires a paired, model-specific neural transceiver in place of an explicit bit interface \cite{ying2025joint}. Across heterogeneous devices and continuously evolving models, such agreements are fragile. The difficulty, we argue, lies not in the token abstraction but in where tokens sit in the stack. This article keeps the tokens and moves them: out of the payload, where they burden the interface, and into the control path, where they guide it.

We call this the {\it token-bit interface transition} and develop it in three steps: 1) {\it unit alignment}, which associates task-side tokens with the corresponding coding units; 2) {\it control derivation}, which converts the aligned relevance scores into codec-specific controls; and 3) {\it priority-preserving coding}, which applies those controls under a single token-derived priority order across both source and channel coding. For image classification, we instantiate the transition as {\it Attention JPEG} with unequal-error-protection (UEP) polar coding, driven by attention scores from a pretrained vision transformer (ViT). Over-the-air experiments on a software-defined radio (SDR) testbed show substantial accuracy--latency gains over conventional SSCC, accuracy competitive with far more memory-intensive neural JSCC in the few-millisecond latency regime, and staged, graceful degradation under channel mismatch.

\section{Tokens as Controls, Not as Cargo}
\begin{figure*}
    \centering
    \includegraphics[width=0.7\linewidth]{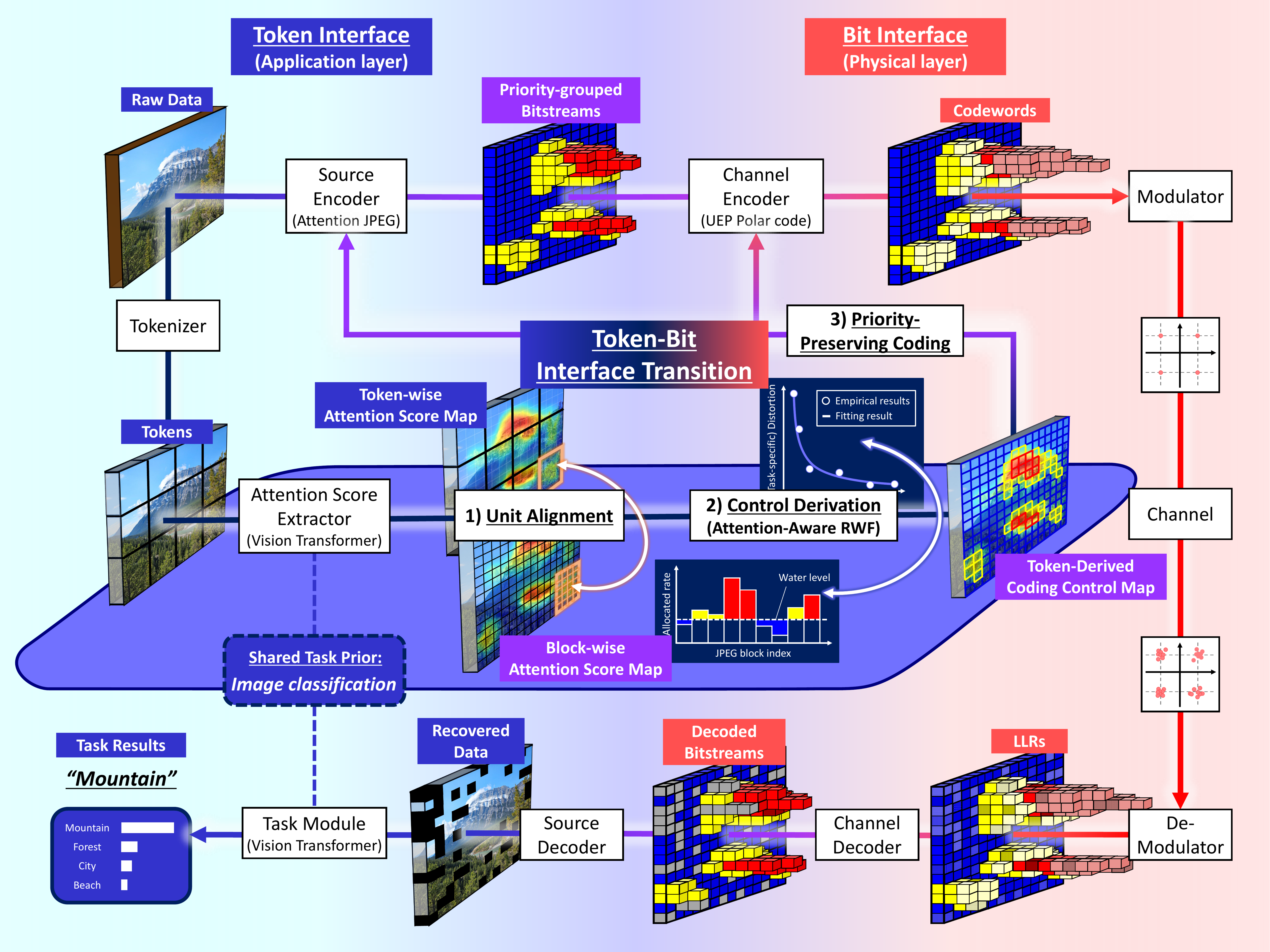}
    \caption{Image-classification instantiation of the token-bit interface transition. The tokenizer (the ViT's patch-embedding stage) defines the task-side token grid, and the attention score extractor (its transformer backbone) computes relevance over that grid. Tokens enter the pipeline only through the coding controls in the shaded region.}
    \label{fig:framework}
\end{figure*}

Consider two endpoints: A holds the source and transmits; B receives and performs a downstream task. We call them endpoints rather than transmitter and receiver because each also hosts an application-layer function---the task-relevance estimator at A, the task module at B---that sits outside the conventional radio boundary and is exactly what the proposed interface couples to the coding chain. Throughout, we assume the source itself is required at B---for archival, human inspection, or tasks beyond the one guiding compression---so A cannot simply transmit its own inference result; the estimator acts as a lightweight proxy for whatever consumes the source at B.

That coding chain is worth keeping. Decades of standardization have equipped the bit interface with interoperable codecs, mature channel codes, retransmission and security machinery, and hardware tuned to digital modulation; any proposal that discards this ecosystem must rebuild it. Our transition is designed so that nothing below the coding controls changes. The same machinery can even amplify the guidance: because the priorities are explicit labels rather than weights inside a network, hybrid ARQ can spend retransmissions on high-priority groups first, and schedulers can order packets accordingly.

Two units must be kept distinct. A {\it token}, as in the token communication literature \cite{devoto2026adaptive, qiao2025token}, is the elementary unit into which an AI model decomposes a source---an image patch, a text sub-word, a short audio segment---and over which it models dependencies to perform inference. What defines it is that partition into inference units, not whether it is stored as a codebook index or a continuous embedding. We say {\it task-side token} to make ownership explicit: the partition belongs to the task models at A and B, and exists whether or not the communication system ever transports it. For each such token, the estimator at A produces a relevance score. A {\it coding unit}, by contrast, is what the communication pipeline compresses and protects: an $8\times 8$ JPEG block, a video slice, a channel-code block. A $224\times 224$ image becomes $196$ tokens of $16\times 16$ pixels under a typical ViT but $784$ JPEG blocks of $8\times 8$ pixels---four blocks per token, with neither partition aware of the other. Reconciling that mismatch is where the interface begins.

Why is this still token communication? Existing designs place tokens in the payload: indices, pruned subsets, or embeddings are transmitted, and the receiver rebuilds the task representation from what arrives \cite{devoto2026adaptive, qiao2025token, ying2025joint}. We keep the identical abstraction and move it to the control path. Tokens still define the units of meaning, token relevance still decides how many bits each part of the source deserves, and the token grid still dictates how the stream is partitioned, prioritized, protected, and allowed to degrade; only the transported quantity changes. The same grounding separates the framework from classical region-of-interest and saliency-guided coding---a lineage running from JPEG2000 ROI coding \cite{skodras2001jpeg} to UEP for progressive image transmission \cite{sherwood1997progressive}---whose mechanics it shares: here the priorities come from the task model's own inference units rather than a hand-crafted perceptual measure, so they track the task automatically and extend to any modality the model tokenizes.

The dividend of the relocation is interoperability. Payload decodability is decoupled from the task model: B needs the transmitted controls, such as a block-wise quality map, but needs neither tokenizer nor token model to decode the payload, so endpoints running different, updated, or no task models still recover the bitstream. Task execution, of course, still requires a task module at B, aligned with A's estimator through a shared task prior.

{\bf 1) Unit alignment} associates each task-side token with the coding unit, or group of units, that carries its source information. The association may be one-to-one, one-to-many, or many-to-one: an image token may span several transform blocks, while several token scores may merge when the codec operates at a coarser resolution. The relevant units may even differ between the source- and channel-coding stages, provided the associations remain traceable as the data are re-represented and regrouped. The aim is not to impose token structure on the codecs, but to decide which units inherit each token's score.

{\bf 2) Control derivation} converts the aligned scores into parameters the codecs actually accept---rates, quality levels, protection levels---in light of the available resources, the codecs' operating characteristics, and the channel. A relevance score says how much a token matters, not what to do about it, so the same score pattern yields different configurations under different constraints: a generous rate budget lets even low-relevance tokens retain moderate quality, while a tight one concentrates rate on a few critical tokens. The mapping is a resource allocation problem, not a lookup table.

{\bf 3) Priority-preserving coding} applies those controls so that a single token-derived priority ordering survives both stages, even when their units and groupings differ---more rate or milder quantization at the source stage, stronger protection at the channel stage. Preserving the ordering keeps the stages from working at cross purposes, heavily protecting content the source coder has already degraded or leaving finely coded content exposed, and steers both compression distortion and residual channel errors toward what the task can spare.

The template is not image-specific. For a language model, tokens are sub-words and the units are text segments: a retrieval system streaming documents to an edge assistant could code the attended passages lightly and the boilerplate aggressively. For speech, tokens are acoustic segments and the units are codec frames, with relevance separating the content a recognizer needs from silence \cite{tian2026large}. For video, tokens tile space and time and the units are slices within a group of pictures. In each case the steps are the same---align, derive, preserve---and only the control parameters change. Porting is not free of labor, since text codecs offer no graceful per-unit quality control and video rate control is entangled with temporal prediction; but that labor lies in rate--distortion (RD) characterization, not in redesigning the interface.

We call the SSCC realization of these steps {\it token-guided SSCC (TG-SSCC)}. Its standard compatibility is architectural rather than bitstream-level: TG-SSCC reuses standardized coding operations---JPEG's transform, quantization, and entropy coding; 5G polar codes; QAM---but its block-wise quality map, DC-predictor resets, and priority grouping require a correspondingly configured decoder. An unmodified, off-the-shelf JPEG receiver will not decode the stream; a lightly reconfigured one, built entirely from standard components, will. And because the task model supplies guidance rather than a task-specific representation, a single pipeline serves many tasks through its control parameters, with no neural transceiver pair trained per task.

In protocol terms, then, the transition is a thin cross-layer interface: the application layer exports a priority map, the lower layers export their control parameters, and neither needs to understand the other's internals. What standardization must supply is correspondingly modest---a format for the map, a channel to carry it, and a way for endpoints to agree on the task prior.

\section{From Attention to Bits: An Image-Classification Example}
\label{sec:imageclass}
We now make the transition concrete for image classification, taking the three steps in order; Fig.~\ref{fig:framework} summarizes the realization. The components---a ViT, JPEG, polar codes---are illustrative, and the procedure carries over to other models and codecs.

\subsection{Step 1: Where Tokens Meet Blocks}
At endpoint A, the patch-embedding stage of a pretrained ViT partitions the image into a fixed grid of patches and projects each into a token embedding. This stage defines the task-side token grid, and hence the spatial support of every relevance score, but produces no relevance itself; the transformer backbone that follows supplies it. The two are successive stages of the same frozen ViT, drawn separately in Fig.~\ref{fig:framework} to distinguish where the grid is defined from where the scores arise.

Within the backbone, attention rollout \cite{abnar2020quantifying} condenses layer-by-layer self-attention into a single normalized score per token: each layer's head-averaged attention matrix is corrected for the residual connection by adding the identity, the matrices are multiplied across layers, and the row of the classification token yields a score for every image token---a direct trace of how much each token influences the classification embedding. We take this as a practical proxy for task relevance. It costs one forward pass through the frozen model, with no additional training and a fraction of the footprint of a learned transceiver (Table~\ref{tab:framework_comparison}). Rollout is one choice among several---gradient-based saliency or a learned relevance head would slot into the same interface---but it needs neither labels nor a backward pass.

Token scores cannot drive JPEG directly, because tokens and JPEG blocks differ in granularity. We assign each token's score to the blocks covering its region, interpolating or aggregating as needed; the resulting block-wise attention map is the input to Step 2.

\subsection{Step 2: Attention-Aware Reverse Water-Filling}
The block-wise map ranks regions; it does not yet allocate rate. Offline, we sweep the JPEG quality levels, measure each level's bit-rate and classification degradation on representative data, and fit low-parameter curves: bit-rate versus quality follows a hyperbolic law and degradation a power law, with separate exponents for the normal and down-sampled representations. Composing the two gives a closed-form RD curve per token.

Online, each token's normalized attention score weights its RD curve, and minimizing the aggregate weighted distortion under the rate budget reduces to a one-dimensional water-level search whose per-token allocation has a closed form. Each allocation is inverted through the rate-quality law and rounded to the nearest supported quality level. The procedure costs a few arithmetic operations per token, runs once per image, involves no training, and leaves nothing to tune by hand: the rate budget goes in, the quality map comes out. When the deployment distribution drifts, the curves are simply re-fit; they live entirely in the control path and touch neither codec nor channel code.

The name reflects the mechanism. Classical reverse water-filling equalizes marginal distortion across sources, and sources already below the common water level receive no rate at all. The attention weights tilt that level token by token---raising it where the classifier is looking, lowering it where it is not---so that equal marginal {\it task} damage, rather than equal marginal mean-squared error, governs the allocation. Quantized to block-wise JPEG quality levels, the resulting allocations form the quality map used by the source coder.

\subsection{Step 3 at the Source: Attention JPEG}
We build on JPEG because it is everywhere: hardware codecs, mature toolchains, three decades of deployed infrastructure. A token interface that upgrades JPEG can upgrade any codec with controllable quantization. JPEG ordinarily applies one quality setting to the entire image; Attention JPEG applies the quality map instead, quantizing each block according to the attention-derived priority of its region while keeping JPEG's transform, quantization, and entropy coding intact. Only the granularity of rate control changes: from image-wide to block-wise.

Two refinements sharpen the instrument. First, fine spatial detail in low-priority regions rarely helps classification, so the blocks of such a token may be down-sampled into a single JPEG block; the two representations carry separate RD curves, and their crossover selects the more rate-efficient option at any allocation. Other coding options can join the same way, by contributing their RD behavior. Second, JPEG's differential coding of direct current (DC) coefficients chains blocks along the raster scan, so one corrupted DC difference can propagate across priority boundaries. We therefore reset the DC predictor at token-aligned boundaries, confining propagation to the region where the error occurs.

\subsection{Step 3 at the Channel: Unequal Error Protection}
Uniform channel protection spends redundancy where the task gains nothing, and can still lose what the task needs. The UEP stage instead carries the same priorities into channel coding. Source-coded blocks are gathered by quality-map tertiles into three priority groups---a grouping B rederives from the quality map itself, with no separate signaling. Each group is encoded by a polar code at its own rate and decoded with cyclic redundancy check (CRC)-aided successive-cancellation list decoding: low rates and generous redundancy for the high-priority groups, high rates for the rest. Polar codes suit this role, since rate is set by the number of frozen bits, making per-group rates a configuration choice rather than a new code design. The number of groups trades protection granularity against signaling and code-block overhead; three sufficed here.

The quality map itself rides in the most strongly protected group. Losing it means losing the parse of the entire payload---the single point whose failure is not graceful---so its protection budget is generous. That generosity costs little: the map is one quality index per token, $196$ entries of a few bits each---on the order of a hundred bytes against a multi-kilobyte payload---and it scales linearly with the token count, not the pixel count. When the channel worsens, low-priority groups fail first: graceful degradation by design, inside a fully digital pipeline.

\section{Over-the-Air Validation}
We implement the framework of Section~\ref{sec:imageclass} end-to-end on an SDR testbed and compare classification accuracy against communication latency, under both matched channel conditions and channel mismatch. A live version of this prototype---the full over-the-air pipeline running semantic-aware rate control in real time---was demonstrated at IEEE ICC 2026 \cite{lee2026semantic}.

\subsection{Testbed, Baselines, and Protocol}
\begin{table*}[t]
\centering
\caption{Implementation details of the evaluated frameworks. A fifth scheme, Attention JPEG with uniform polar coding, matches the TG-SSCC column with UEP disabled.}
\label{tab:framework_comparison}
\renewcommand{\arraystretch}{1.1}
\setlength{\tabcolsep}{3pt}
\begin{tabularx}{\textwidth}{c|c|*{4}{>{\centering\arraybackslash}X|}}
\hline
\multicolumn{2}{c|}{}
& \textbf{SSCC} & \textbf{TG-SSCC} & \textbf{JSCC} & \textbf{TG-JSCC} \\
\hline\hline
\multicolumn{2}{c|}{\textbf{Implementation}}
& JPEG + 5G standard polar
& \makecell{Attention JPEG +\\5G standard polar with UEP}
& NTSCC
& \makecell{NTSCC + attention-aware\\symbol-length allocation} \\
\hline
\multicolumn{2}{c|}{\textbf{Payload}}
& \multicolumn{2}{c|}{DC/AC bit-stream for Y/Cb/Cr channels}
& \multicolumn{2}{c|}{Continuous-valued symbol stream} \\
\hline
\multicolumn{2}{c|}{\textbf{Side information}}
& Image-level JPEG quality
& Block-wise quality map
& \multicolumn{2}{c|}{\makecell{Per-token transmit-symbol lengths\\(NTSCC latent tokens)}} \\
\hline
\multicolumn{2}{c|}{\textbf{Transmitter Model Memory}}
& --
& \makecell{DeiT-Tiny: 22\,MB}
& \makecell{NTSCC: 132\,MB}
& \makecell{NTSCC: 132\,MB\\ \& DeiT-Tiny: 22\,MB} \\
\hline
\multirow{2}{*}{\textbf{Constellation}}
& \textbf{Payload}
& Digital (QAM)
& Digital (QAM)
& Analog
& Analog \\
\cline{2-6}
& \textbf{Side info.}
& Digital (QAM)
& Digital (QAM)
& Digital (QAM)
& Digital (QAM) \\
\hline
\end{tabularx}
\end{table*}
\begin{figure}[t]
    \centering
    \includegraphics[width=0.95\columnwidth]{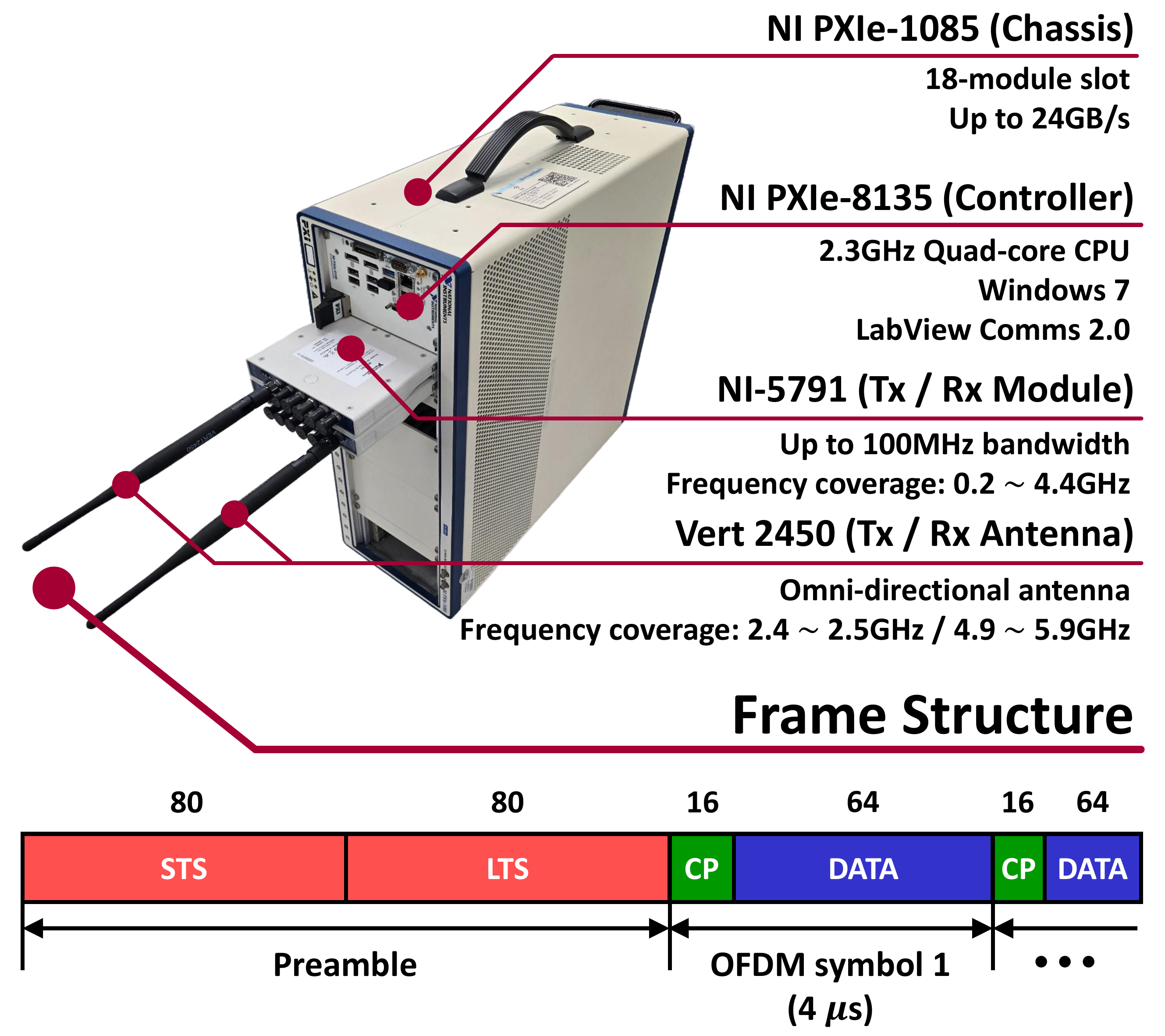}
    \caption{NI PXIe-based over-the-air testbed: hardware configuration and OFDM frame structure.}
\label{fig:equip}
\end{figure}

We evaluate on a 100-class subset of ImageNet \cite{deng2009imagenet}, 30 images per class. A frozen DeiT-Tiny model \cite{touvron2021training} supplies attention scores at A and classifies the reconstructed images at B; it is deliberately small, testing whether a lightweight frozen model suffices to guide coding, and because it also performs the classification, its capacity caps the achievable accuracy. The figure of merit is top-1 accuracy versus orthogonal frequency division multiplexing (OFDM) transmission latency, referenced to accuracy on the original images. Latency counts the airtime of everything a scheme transmits, payload and side information alike, and excludes computation at either endpoint, which we do not optimize here.

Table~\ref{tab:framework_comparison} details the four frameworks. Conventional SSCC pairs JPEG with 5G-standard polar coding; TG-SSCC pairs Attention JPEG with UEP polar coding over three priority groups. To isolate the source-coding contribution, we additionally run Attention JPEG with the same uniform CRC-aided polar code as conventional SSCC. Nonlinear transform source-channel coding (NTSCC) \cite{dai2022nonlinear} serves as the neural JSCC baseline, and TG-JSCC is an NTSCC variant in which attention scores guide per-token symbol-length allocation---token-guided control applied to JSCC. Each scheme digitally conveys the side information B needs to interpret its payload. The memory row of Table~\ref{tab:framework_comparison} concerns the transmitter only; at B, every scheme hosts the same classifier, and the JSCC family additionally its decoder network.

All schemes share the same OFDM frame, synchronization, and equalization on the NI PXIe-based testbed of Fig.~\ref{fig:equip}, differing only in what fills the payload symbols---QAM for the SSCC family, continuous-valued symbols for the JSCC family---so the comparison isolates the coding chain rather than the radio. Under matched operation, configurations are selected for the operating signal-to-noise ratio (SNR); under mismatch, they are held fixed while the actual SNR varies.

\subsection{When the Channel Is Known: Matched Operation}
\begin{figure*}[t]
    \centerline{
    \hspace*{\fill}
    \subfloat[SNR-Matched (SNR: 3 dB)]{%
        \includegraphics[width=0.57\columnwidth]{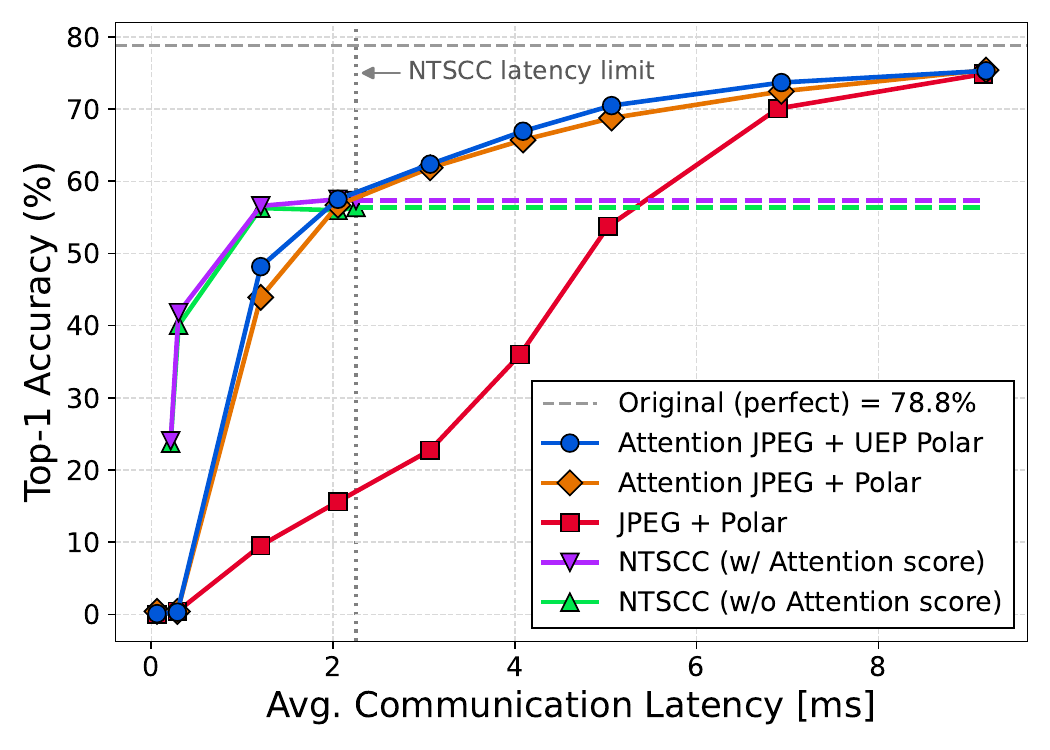}}
    \hfill
    \subfloat[SNR-Mismatched (SNR: 1.5 dB)]{%
        \includegraphics[width=0.57\columnwidth]{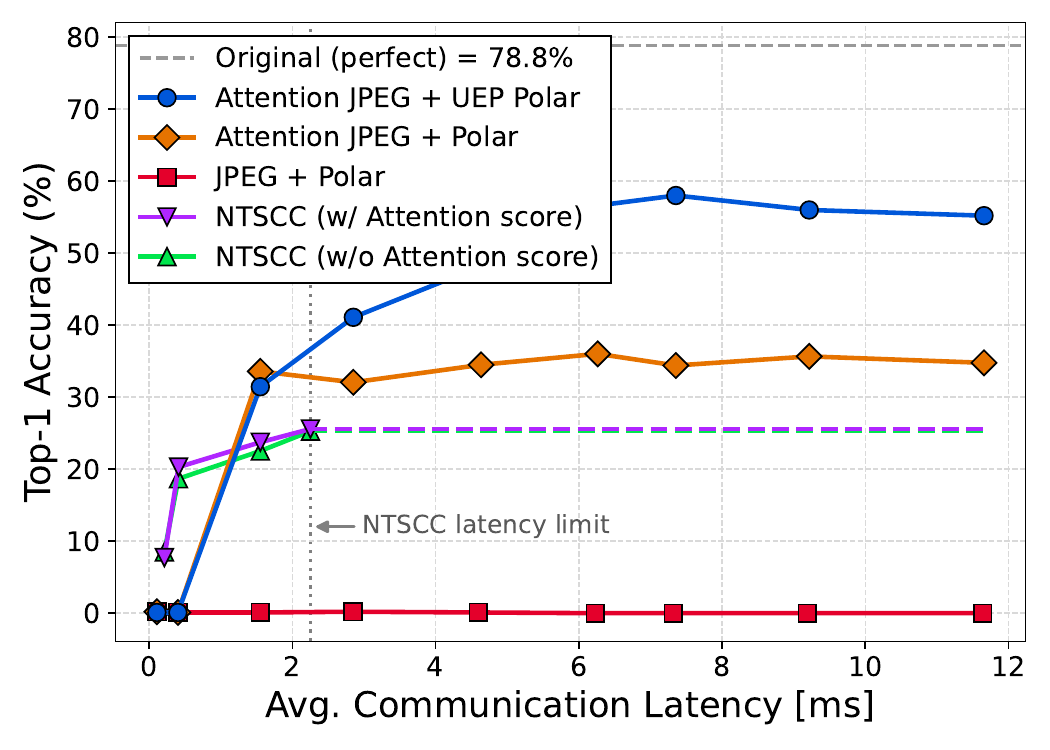}}
    \hfill
    \subfloat[SNR-Mismatched (Latency: 7.44 ms)]{%
        \includegraphics[width=0.57\columnwidth]{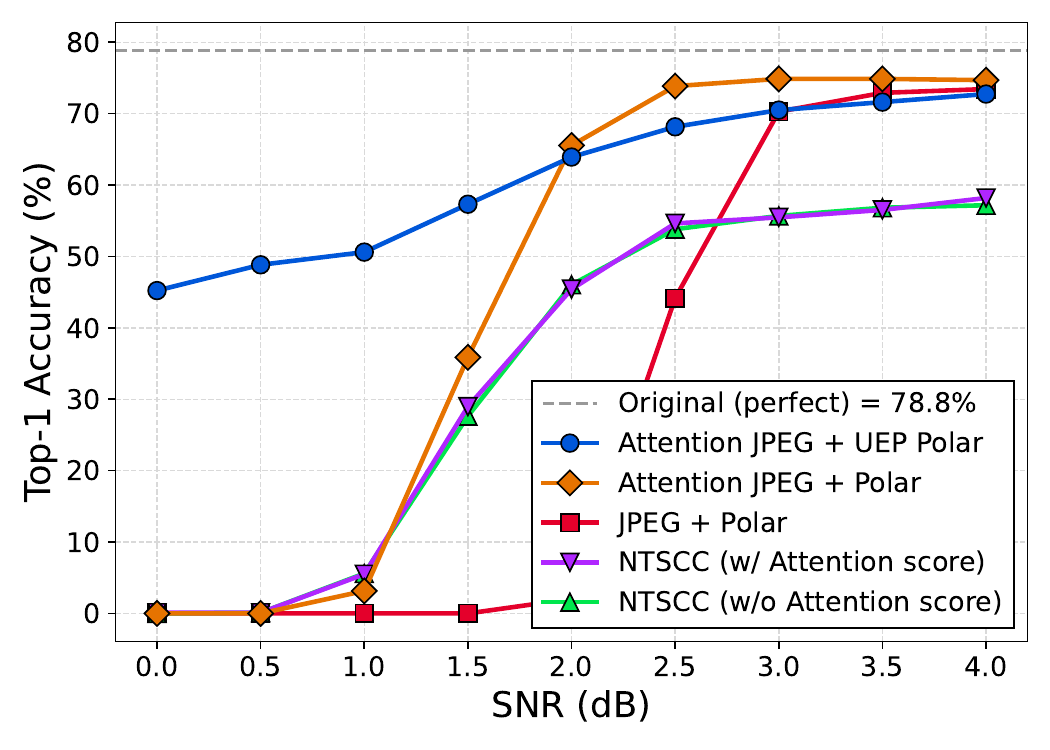}}
    \hspace*{\fill}
    }
    \caption{Top-1 classification performance under SNR-matched and SNR-mismatched operation. Under matched operation, the coding and model configurations are selected for the operating SNR. Under mismatched operation, uniform polar coding uses the fixed code rate $R=0.5$, whereas UEP uses $R_1=0.25$, $R_2=0.5$, and $R_3=0.75$ for the high-, medium-, and low-priority groups, respectively. In (a) and (b), the curves labeled NTSCC with/without attention scores correspond to TG-JSCC/NTSCC in the text; both end at the maximum latency the trained model supports, since its architecture admits only a bounded range of transmit-symbol lengths. Horizontal dashed segments extend each terminal NTSCC accuracy as a visual reference only.}
\label{fig:acc}
\end{figure*}

Fig.~\ref{fig:acc}(a) reports matched operation at 3 dB. The gap between {\it JPEG + Polar} and {\it Attention JPEG + Polar} isolates the source-coding effect, since both use the same matched polar code with negligible residual decoding errors. To reach a given latency, conventional JPEG must compress the entire image aggressively, distorting the object region along with everything else; Attention JPEG spends the same bits unevenly and preserves what the classifier looks at, as Fig.~\ref{fig:imgs}(b) makes visible. The complete TG-SSCC, {\it Attention JPEG + UEP Polar}, then carries those priorities into channel coding, raising code rates on low-priority groups while keeping task-relevant content strongly protected. The intermediate variant thus separates the interface's two contributions: its gain over conventional SSCC is the source-side dividend, whereas the further gain of UEP is modest here---the channel-side dividend matures under mismatch.

The JSCC baselines behave differently. {\it NTSCC}, trained separately for each operating SNR, excels in the ultra-low-latency regime, where its learned nonlinear transform extracts a compact joint representation that no fixed transform matches at extreme compression---at the price of substantially greater model memory and computation. That comparison should be read with its purpose in mind: NTSCC learns a general reconstruction codec, a harder objective than guiding a classifier. Its architecture also supports only a bounded range of transmit-symbol lengths, so its curve ends at the largest latency the trained model can produce, not at a converged accuracy; extending that range means retraining a larger model, not adjusting a run-time parameter. We therefore compare quantitatively only over the latency region all schemes reach, and claim nothing about NTSCC beyond its endpoint. The contrast we draw is one of operating range and reconfigurability: TG-SSCC spans the full latency axis with one fixed pipeline and roughly one-sixth the transmitter model memory. TG-JSCC adds a further gain over NTSCC through token-wise symbol-length allocation---evidence that token-guided control is not confined to SSCC.

The axis is latency rather than SNR because, under matched operation, an SNR axis is uninformative: every scheme reconfigures for the operating SNR---NTSCC is even retrained---so each abscissa would compare a different configuration with negligible residual errors at its own SNR. The operating SNR changes only the latency needed to deliver the same content; the SNR axis earns its place once the configuration is frozen and the channel drifts.

\subsection{When It Is Not: Graceful Degradation under Mismatch}
\begin{figure*}[t]
\centering
\subfloat[Original]{%
    \begin{minipage}[b]{0.33\columnwidth}
        \centering
        \includegraphics[width=\linewidth]{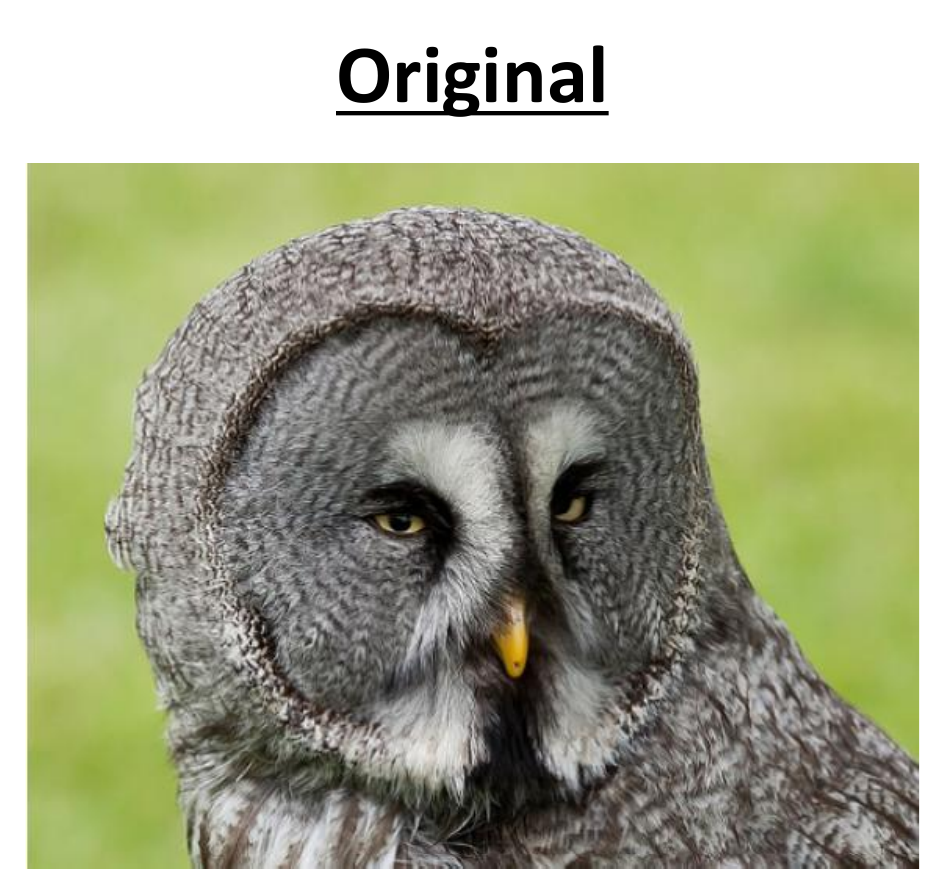}
        \vspace{0.12\columnwidth}
    \end{minipage}}
\hfill
\subfloat[SNR-Matched (SNR: 3 dB, latency: 3.65 ms)]{%
    \includegraphics[width=0.66\columnwidth]{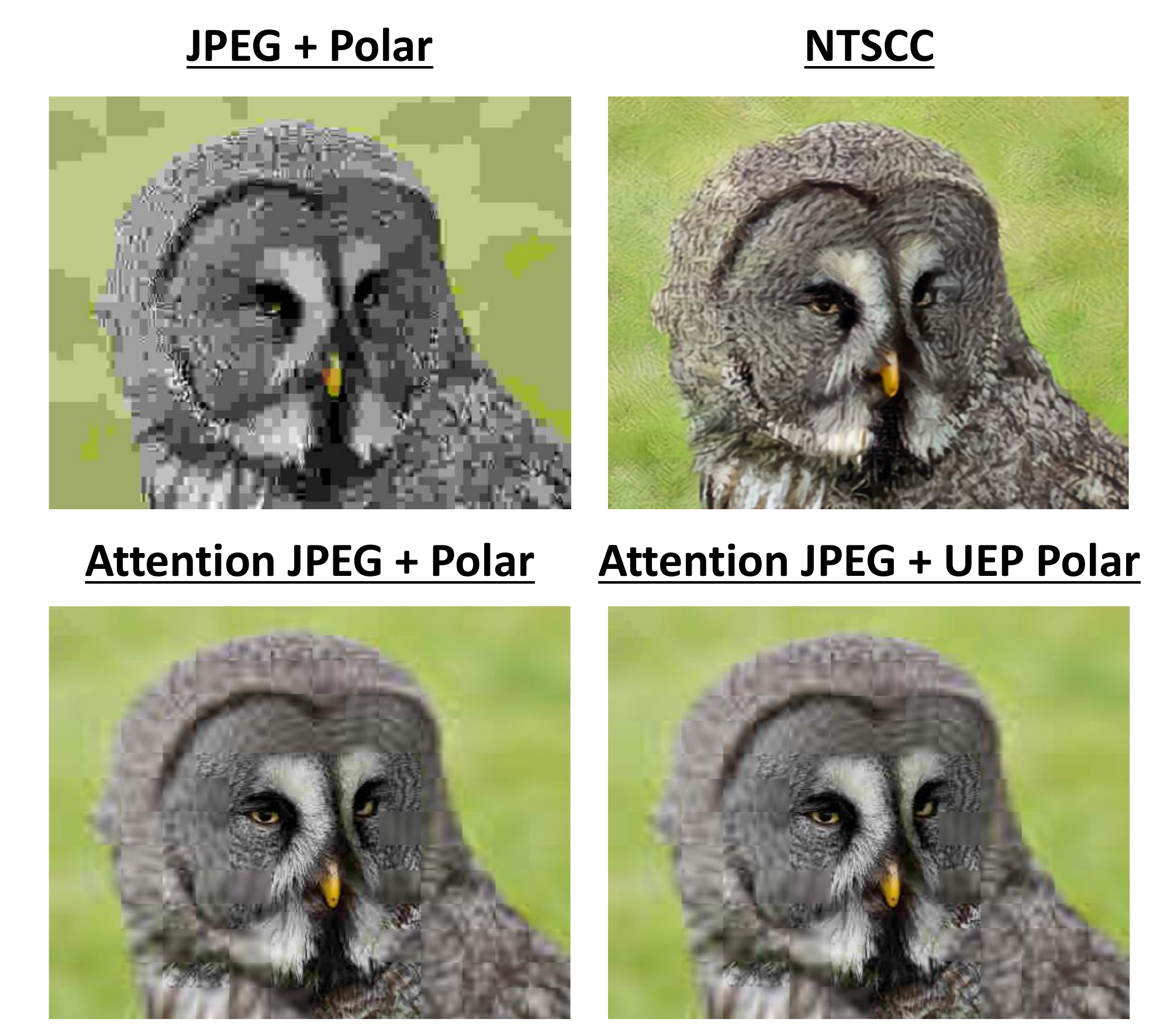}}
\hfill
\subfloat[SNR-Mismatched (SNR: 1.5 dB, latency: 6.91 ms)]{%
    \includegraphics[width=0.66\columnwidth]{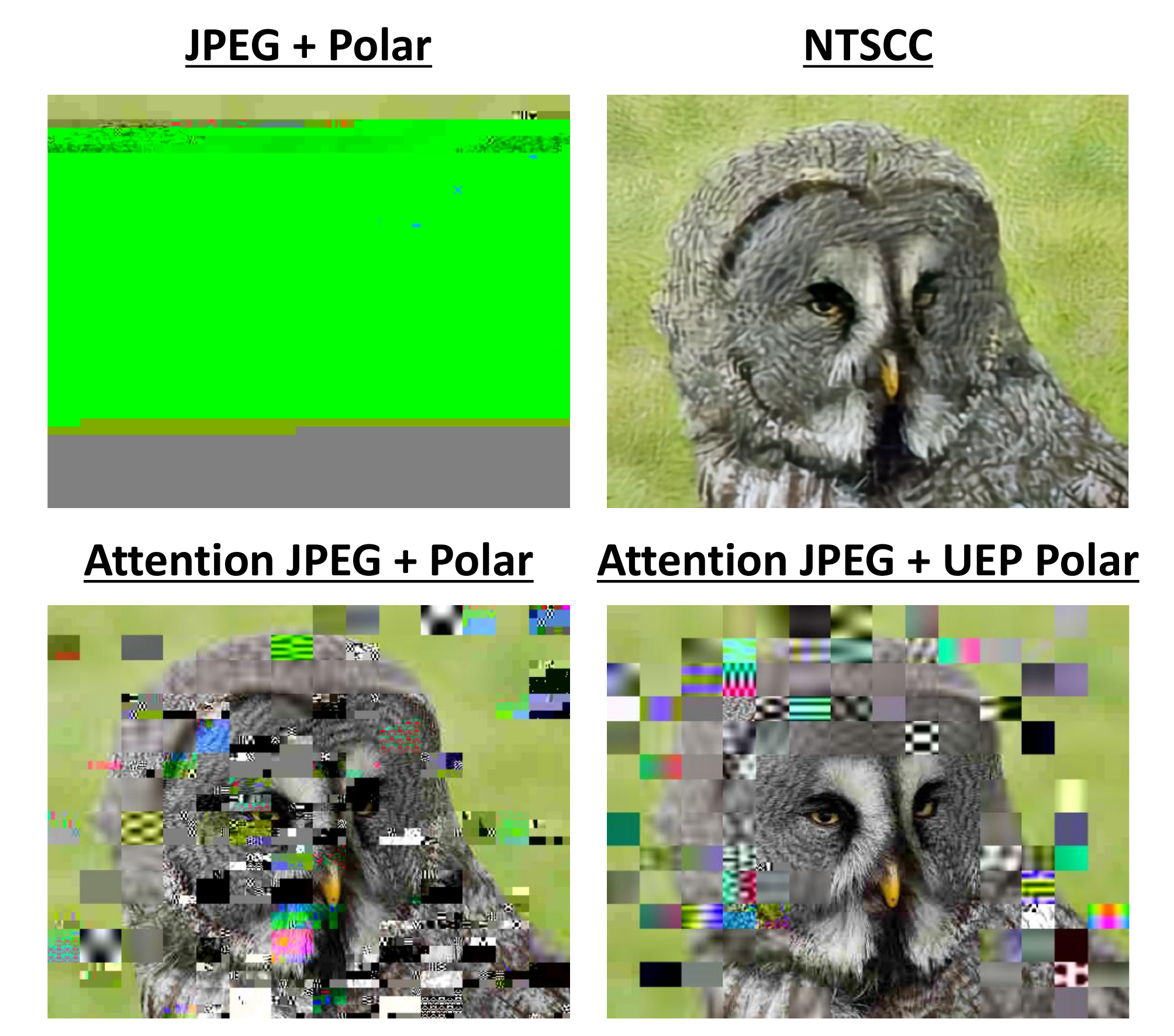}}
\caption{Representative reconstruction results for one test image under SNR-matched and SNR-mismatched operation. In (c), visual quality and task success diverge: TG-SSCC, though visibly corrupted in low-priority regions, preserves the object and is classified correctly; the uniformly protected digital schemes are not.}
\label{fig:imgs}
\end{figure*}

Under mismatch, both endpoints retain configurations chosen for 3 dB while the channel does not cooperate---a setting that probes what the transition buys when channel knowledge fails, as SNR estimates age quickly in practice. At 1.5 dB, {\it JPEG + Polar} collapses: decoding errors strike, and image-wide DC prediction drags the corruption across the raster scan, producing the complete reconstruction failure of Fig.~\ref{fig:imgs}(c). {\it Attention JPEG + Polar} confines errors to individual token regions, but its uniform protection ignores the priorities at the bit interface, so object regions fail as readily as background. TG-SSCC instead lets low-priority groups fail while the high-priority ones---the object regions and the quality map---remain decodable and support correct classification. NTSCC stays recognizable, its continuous-valued payload having no digital decoding cliff, though its reconstruction quality erodes.

The SNR sweep in Fig.~\ref{fig:acc}(c) summarizes the mechanics. Uniformly protected digital schemes fall off a cliff once their fixed-rate codes can no longer deliver the bitstreams. TG-SSCC maps the token-derived ordering onto explicit bit-level protection, so groups fail in reverse order of importance: staged, graceful degradation within a fully digital pipeline, with the quality map and high-priority regions surviving over a wider SNR range. NTSCC degrades gently as long as its digital side information---the per-token symbol lengths---remains decodable, then loses everything at once when it does not. The hedge has a price, visible in the same figure: at and above the design SNR, uniform protection outperforms UEP by several points, redundancy spent on insurance that was not needed.

\subsection{Lessons and Limitations}
Three lessons emerge, each tied to a claim of Section II. First, much of the gain over conventional SSCC arrives at the source stage: aligning compression with the task's attention pattern is cheap and requires no change to channel coding. Second, the channel-side dividend appears exactly when the channel misbehaves---under matched conditions UEP mainly buys rate flexibility; under mismatch it buys survival. Third, graceful degradation, often cited as the signature advantage of analog JSCC, is not exclusive to analog modulation: a digital pipeline degrades gracefully when its bitstream is partitioned along priorities and protected accordingly, which is what the token grid provides.

A quieter fourth lesson concerns side information. Every scheme evaluated here, neural ones included, depends on a small digital control stream to interpret its payload, and every abrupt failure we observed traces back to losing it. Whatever else changes, the control plane comes first.

The evaluation has limits worth stating plainly. It covers one task, one task model, and one codec pair; A and B share that model, so cross-model transfer of the attention guidance remains untested; attention rollout is a heuristic proxy whose agreement with ground-truth importance is not guaranteed; accuracies are averaged without confidence intervals from repeated runs; and the relevance signal remains to be ablated over the air against saliency-based and random baselines. Closing these gaps is routine work, and none of it alters the architectural point: the interface exists, runs over the air, and requires nothing below the coding controls to change.

\section{Conclusion: How Far Can Task Awareness Go?}
This article proposed treating tokens as the controllers of a digital communication pipeline rather than as its cargo. The token-bit interface transition aligns task-side tokens with coding units, derives codec-specific controls from their relevance, and preserves the resulting priorities across source and channel coding. Attention JPEG and UEP polar coding realize it for image classification, and over-the-air experiments confirm substantial gains over conventional SSCC, accuracy competitive with far more memory-intensive neural JSCC, and graceful degradation under mismatch.

Each step invites generalization. Unit alignment should span more modalities and token structures; control derivation should couple relevance extraction to RD characterization, so that controls reflect the true task benefit of each coding unit; priority-preserving coding should push task adaptation deeper into the codecs without surrendering the modularity, decodability, and signaling discipline of the bit interface. A priority map sent in the clear also tells an eavesdropper which parts of a transmission matter, so control-plane confidentiality deserves study alongside robustness. How far task awareness can go before the virtues of the bit interface erode is the central open question.

The timing matters: 6G standardization is beginning while AI models change by the month, so an interface that survives model churn is worth more than one tied to a frozen model. Fully learned transceivers mark one endpoint of that evolution, but their model coupling hinders near-term deployment, and standardization moves faster on control interfaces than on new payloads: signaling a quality map is a far smaller step than specifying a neural transceiver. The transition proposed here lets today's bit-oriented infrastructure start thinking in tokens---while still talking in bits.

\bibliographystyle{IEEEtran}
\bibliography{bibfile_2}

@article{skodras2001jpeg,
    author = {Skodras, Athanassios and Christopoulos, Charilaos and Ebrahimi, Touradj},
    title = {The {JPEG} 2000 still image compression standard},
    journal = {IEEE Signal Process. Mag.},
    year = {2001},
    volume = {18},
    number = {5},
    pages = {36--58}
}

@article{sherwood1997progressive,
    author = {Sherwood, P. Greg and Zeger, Kenneth},
    title = {Progressive image coding for noisy channels},
    journal = {IEEE Signal Process. Lett.},
    year = {1997},
    volume = {4},
    number = {7},
    pages = {189--191}
}

@article{gunduz2023beyond,
    author = {G{\"u}nd{\"u}z, Deniz and Qin, Zhijin and Aguerri, Inaki Estella and Dhillon, Harpreet S and Yang, Zhaohui and Yener, Aylin and Wong, Kai Kit and Chae, Chan-Byoung},
    title = {Beyond transmitting bits: Context, semantics, and task-oriented communications},
    journal = {IEEE J. Sel. Areas Commun.},
    year = {2023},
    volume = {41},
    number = {1},
    pages = {5--41}
}

@article{devoto2026adaptive,
    author = {Devoto, Alessio and Pomponi, Jary and Merluzzi, Mattia and Di Lorenzo, Paolo and Scardapane, Simone},
    title = {Adaptive semantic token communication for transformer-based edge inference},
    journal = {IEEE Trans. Mach. Learn. Commun. Netw.},
    year = {2026},
    volume = {4},
    pages = {422--437}
}

@article{qiao2025token,
    author = {Qiao, Li and Mashhadi, Mahdi Boloursaz and Gao, Zhen and Tafazolli, Rahim and Bennis, Mehdi and Niyato, Dusit},
    title = {Token communications: A large model-driven framework for cross-modal context-aware semantic communications},
    journal = {IEEE Wireless Commun.},
    year = {2025},
    volume = {32},
    number = {5},
    pages = {80--88}
}

@article{ying2025joint,
    author = {Ying, Jingkai and Qin, Zhijin and Feng, Yulong and Wang, Liejun and Tao, Xiaoming},
    title = {Joint semantic-channel coding and modulation for token communications},
    journal = {IEEE Trans. Wireless Commun.},
    year = {2026},
    volume = {25},
    pages = {8179--8193}
}

@article{bourtsoulatze2019deep,
    author = {Bourtsoulatze, Eirina and Kurka, David Burth and G{\"u}nd{\"u}z, Deniz},
    title = {Deep joint source-channel coding for wireless image transmission},
    journal = {IEEE Trans. Cogn. Commun. Netw.},
    year = {2019},
    volume = {5},
    number = {3},
    pages = {567--579}
}

@article{lyu2024semantic,
    author = {Lyu, Zhonghao and Zhu, Guangxu and Xu, Jie and Ai, Bo and Cui, Shuguang},
    title = {Semantic communications for image recovery and classification via deep joint source and channel coding},
    journal = {IEEE Trans. Wireless Commun.},
    year = {2024},
    volume = {23},
    number = {8},
    pages = {8388--8404}
}

@article{huang2025d2,
    author = {Huang, Jianhao and Yuan, Kai and Huang, Chuan and Huang, Kaibin},
    title = {D$^2$-{JSCC}: Digital deep joint source-channel coding for semantic communications},
    journal = {IEEE J. Sel. Areas Commun.},
    year = {2025},
    volume = {43},
    number = {4},
    pages = {1246--1261}
}

@article{dai2022nonlinear,
    author = {Dai, Jincheng and Wang, Sixian and Tan, Kailin and Si, Zhongwei and Qin, Xiaoqi and Niu, Kai and Zhang, Ping},
    title = {Nonlinear transform source-channel coding for semantic communications},
    journal = {IEEE J. Sel. Areas Commun.},
    year = {2022},
    volume = {40},
    number = {8},
    pages = {2300--2316}
}

@article{tian2026large,
    author = {Tian, Yun and Qin, Zhijin and Lv, Guocheng and Jin, Ye and Huang, Kaibin and Han, Zhu},
    title = {Large speech model enabled semantic communication},
    journal = {IEEE Trans. Mobile Comput.},
    year = {2026},
    volume = {25},
    number = {8},
    pages = {11600--11614}
}

@inproceedings{abnar2020quantifying,
    author = {Abnar, Samira and Zuidema, Willem},
    title = {Quantifying attention flow in transformers},
    booktitle = {Proc. 58th Annu. Meeting Assoc. Comput. Linguistics (ACL)},
    year = {2020}
}

@inproceedings{deng2009imagenet,
    author = {Deng, Jia and Dong, Wei and Socher, Richard and Li, Li-Jia and Li, Kai and Fei-Fei, Li},
    title = {{ImageNet}: A large-scale hierarchical image database},
    booktitle = {Proc. IEEE Conf. Comput. Vis. Pattern Recognit. (CVPR)},
    year = {2009}
}

@inproceedings{touvron2021training,
    author = {Touvron, Hugo and Cord, Matthieu and Douze, Matthijs and Massa, Francisco and Sablayrolles, Alexandre and J{\'e}gou, Herv{\'e}},
    title = {Training data-efficient image transformers \& distillation through attention},
    booktitle = {Proc. Int. Conf. Mach. Learn. (ICML)},
    year = {2021}
}

@misc{lee2026semantic,
    author = {Lee, Namyoon and Park, Chanho},
    title = {Semantic-aware rate control for wireless communication: A prototype demonstration},
    howpublished = {Industry Demo, IEEE Int. Conf. Commun. (ICC), Glasgow, U.K.},
    month = may,
    year = {2026},
    note = {[Online]. Available: \url{https://icc2026.ieee-icc.org/program/industry-demos}}
}

\begin{IEEEbiographynophoto}{Chanho Park}
(Member, IEEE) received the Ph. D. degrees from POSTECH in 2026 in electrical engineering. He is currently a postdoctoral researcher with POSTECH, Pohang, South Korea. His research interests include AI-native communication and channel coding.
\end{IEEEbiographynophoto}

\begin{IEEEbiographynophoto}{Bumsu Park}
(Student Member, IEEE) is currently pursuing the Ph.D. degree in electrical engineering with POSTECH, Pohang, South Korea. His research interests include task-oriented source and channel coding and SDR implementation of semantic communication systems.
\end{IEEEbiographynophoto}

\begin{IEEEbiographynophoto}{Soonhee Kwon}
received the Ph. D. degrees from Hanyang University in 2023. He is currently a Senior Research Engineer with the Communication \& Media Standard Laboratory, LG Electronics, Seoul, South Korea. Her research interests include semantic communication and 6G air-interface design.
\end{IEEEbiographynophoto}

\begin{IEEEbiographynophoto}{Sangrim Lee}
(Member, IEEE) received the Ph.D. degree from Korea University in 2013. He is currently a Senior Research Engineer with the Communication \& Media Standard Laboratory, LG Electronics, Seoul, South Korea. His research interests include AI-native wireless communication and 3GPP standardization.
\end{IEEEbiographynophoto}

\begin{IEEEbiographynophoto}{Namyoon Lee}
(Senior Member, IEEE) received the Ph.D. degree from the University of Texas at Austin in 2014. He is currently a Professor with the Department of Electrical Engineering, POSTECH, Pohang, South Korea. His research interests include communication theory, channel coding, and machine learning for wireless systems.
\end{IEEEbiographynophoto}

\end{document}